# Estimating the Effort Overhead in Global Software Development

Ansgar Lamersdorf
*University of Kaiserslautern*
*Kaiserslautern, Germany*
a_lamers@informatik.uni-kl.de

Jürgen Münch
*University of Kaiserslautern and Fraunhofer IESE*
*Kaiserslautern, Germany*
Juergen.Muench@iese.fraunhofer.de

Alicia Fernández-del Viso Torre
*Indra Software Labs*
*Madrid, Spain*
afernandezde@indra.es

Carlos Rebate Sánchez
*Indra Software Labs*
*Madrid, Spain*
crebate@indra.es

Dieter Rombach
*University of Kaiserslautern and Fraunhofer IESE*
*Kaiserslautern, Germany*
Dieter.Rombach@iese.fraunhofer.de

*Abstract*— Models for effort and cost estimation are important for distributed software development as well as for collocated software and system development. Standard cost models only insufficiently consider the characteristics of distributed development such as dissimilar abilities at the different sites or significant overhead due to remote collaboration. Therefore, explicit cost models for distributed development are needed. In this article, we present the initial development of a cost overhead model for a Spanish global software development organization. The model was developed using the CoBRA approach for cost estimation. As a result, cost drivers for the specific distributed development context were identified and their impact was quantified on an empirical basis. The article presents related work, an overview of the approach, and its application in the industrial context. Finally, we sketch the inclusion of the model in an approach for systematic task allocation and give an overview of future work.

## I. INTRODUCTION

The estimation of effort and cost early in a software development project has been in the focus of software engineering research for a long time and has led to the development of several approaches and models for effort estimation [1-5]. However, these models usually focus on typical development projects and do not specifically consider distributed or global software development [6-8]. COCOMO2 [2], for example, includes only one effort multiplier for describing the distribution of work.

On the other hand, it is well known that the distribution of work has a significant impact on productivity and that distributed tasks can take much longer compared to collocated work [9]. Thus, cost analysis and estimation in global software development (GSD) is much more complex than in collocated development [10]. In addition, multiple sites have to be coordinated and are dependent on each other in GSD, which increases the need for reliable estimates of the effort involved at each site.

At the same time, GSD projects often have a high failure rate [11] and, in particular, suffer from cost overruns [12, 13], which indicates that there is very little knowledge about the impact of work distribution on effort in GSD. Therefore, we need new cost estimation models and methods that specifically consider the characteristics of distributed development and are able to provide support for effort prediction in GSD projects.

Cost estimation for GSD differs from standard cost estimation in two dimensions: First, there is a large overhead that is due to the distributed collaboration and impacted by several factors such as language differences [14-16], cultural barriers [14, 15, 17, 18], or time shifts [19, 20] between sites. Second, many effort drivers (such as the ability of the workforce) are site-specific and cannot be considered globally for a project [8]: In many projects, the development sites have very different characteristics and thus the productivity rate is different between sites. A standard effort prediction model consisting only of global effort drivers cannot account for these differences.

A cost estimation method that includes specific cost overheads drivers and site-specific characteristics could also be used for systematic task allocation: In the planning of GSD projects, it has to be decided which sites to include in a project and which work to assign to which site [21]. This decision is very complex as it has to include multiple factors and criteria [22, 23]. If a cost estimation method is able to predict the effort with respect to site-specific characteristics and overhead drivers, it can be used to evaluate different assignment scenarios and thus to make a systematic assignment decision.

In this article, we present the causal modeling and quantification of cost overhead factors in the context of a global software development organization based in Spain. This can be used for the creation of a cost estimation model based on the CoBRA approach, as this was designed specifically for developing individual cost models using expert estimations and project data. The model therefore contains an empirically based description of the most important overhead drivers for distributed collaboration. Due to the scope of the study, it does not specifically address individual effort drivers for each site but is restricted to projects with only one remote site. We will, however, describe how site-specific effort drivers can be included in the model.

In the standard literature on effort estimation, no difference is usually made between *effort* and *costs*. This is mainly caused by the fact that in software development, practically all costs are personnel costs, which are directly dependent on effort. In GSD, on the other hand, cost rates might differ between sites, meaning that effort at one site might cause higher costs than effort at another site. However, we will, in the following, use effort and cost as synonyms – except for Section 5, where we sketch the inclusion of the effort model in a task allocation approach.

The remainder of this article is structured as follows. First, we will discuss related work on cost estimation for GSD. Section 3 presents the CoBRA methodology and its application for the development of the GSD cost model. The model and identified cost drivers are presented in Section 4. In Section 5, we demonstrate how the model can be extended towards including site-specific cost drivers and show its application for task allocation evaluation, followed by a conclusion and a discussion on future work.

## II. RELATED WORK

This section will present related work in effort estimation for software development projects by first giving a rough overview of effort estimation in general and then of specific approaches for GSD effort prediction.

### A. Effort Estimation in Software Engineering

In general, effort estimation approaches aim at predicting the effort for a development project based on its expected size (e.g., in lines of code or function points) and project-specific characteristics. Two types can be distinguished: data-driven and expert-driven approaches [5].

Data-driven approaches try to identify relations between size, effort drivers, and project effort based on the data from past projects. The best-known data-driven approach in the literature is COCOMO(2) [2], where project effort is predicted according to the function

$$Effort = A * Size^B * \prod_{i=1}^{n} EM_i$$

with A and B being constants and EM a set of effort multipliers. The values for A, B, and the effort multipliers were derived from an analysis of a large number of past projects.

The standard COCOMO2 approach does not address the characteristics of distributed development in great detail. There is one effort multiplier for multi-site development but as this is only a single number, it cannot reflect the inherent complexity and various overhead drivers of global software development.

Despite the standard model with its published parameters and effort multipliers, COCOMO can also be seen as a framework for building one's own effort models, which could include GSD-specific influencing factors. However, this would require access to a very large number of projects in order to get quantifications for the parameters via statistical regression. In addition, COCOMO does not provide a methodology for evaluating different sites with different parameter values (e.g., different levels of experience at different sites).

Other data-driven approaches include machine learning approaches [24], optimized set reduction [3], stepwise analysis of variance [25], or ordinary least square regression [26]. For these approaches, no complete models have been published; they only describe ways to develop organization-specific effort models. Therefore, they could be used to create effort estimation models that address the specifics of GSD. However, this would again require a very large data set of past projects, which can not be gathered in most organizations.

In contrast to data-driven methods, expert-based approaches rely on the expertise of experienced practitioners rather than on historical project data. One wide-spread approach is the Delphi method, where several experts are asked to give estimations regarding project effort in several rounds [27]. In agile development, the Planning Game [28] is often used for estimating the effort needed for small increments in a development project. The advantage of these approaches is that no historic data is needed; thus, they are much easier to apply in practice. However, the estimation process is not transparent and relies only on the expertise of the involved persons. In addition, no statement is made about the impact of specific characteristics (e.g., cultural distance) on project effort and thus it cannot be used to plan and optimize future projects.

There exist also hybrid methods that try to combine data-driven and expert-based approaches. One example is the CoBRA (COst estimation, Benchmarking, and Risk Assessment) method [29]. Here, effort drivers are identified and quantified in a systematic process using expert estimation, while a baseline productivity is derived from a small set of projects. This method provides the advantages of data-driven approaches (a systematic, transparent estimation model) while needing only little project data and thus being applicable in most industrial environments. CoBRA has been successfully applied several times in different software development environments [30-32].

### B. Specific Effort Estimation Approaches for GSD

Few approaches to effort estimation exist that specifically address distributed and global software development by suggesting extensions for standard effort prediction approaches. As COCOMO is the most prominent and best-documented approach for effort estimation in the literature, they are all based on COCOMO.

Keil et al. sketch an idea on how to extend COCOMO towards GSD by suggesting a set of new effort multipliers for distributed development [6]. As in the original COCOMO, the multipliers are categorized into product, personnel, and project factors and describe characteristics such as architectural adequacy, cultural fit, and physical distance. The factors were derived from the authors' personal experiences and form other publications. However, they did not appear to have been collected in a systematic process. A quantification of the factors is also not given.

A similar approach is suggested by Betz and Mäkiö [7]. They again identified a set of additional effort multipliers that describe the characteristics of global and outsourced software development. Here, they are classified into outsourcing factors, factors describing the buyers' and providers' outsourcing maturity, and coordination factors. These factors were identified based on qualitative interviews with practitioners and were also quantified in terms of their impact. However, this quantification was not done in a systematic process and it is not clear where the numeric values for the impact of the factors stem from.

Madachy [8] presents a different solution for adapting COCOMO for distributed development. Based on the fact that different sites in a GSD project might have different characteristics, he suggests identifying values for the effort multipliers $EM_i$ individually for every site. Thus, the original COCOMO formula is modified into one formula for calculating the unadjusted effort using the project size and the parameters A and B. This unadjusted effort is then split up and assigned to the different sites where the adjusted effort is calculated according to the site-specific multipliers. The approach addresses the problem of having site-specific characteristics in GSD. However, it does not regard the additional overhead in distributed development that is caused by GSD-specific factors such as cultural or time zone differences.

The analysis of related work shows that existing approaches in the literature recognize the fact that in GSD, there are specific additional cost drivers that have to be considered in effort prediction. However, they do not present systematic estimations or quantification on the impact of these drivers. Therefore, we decided to develop our own, specific effort model for effort estimation in a specific global software development environment. Due to its advantages of resulting in a defined effort model without being dependent on large amounts of project data, we decided to use the CoBRA approach for model development.

III. RESEARCH METHOD

In this section, we will first present the research goal and the main research questions. As our work is mainly based on CoBRA, we will then introduce the CoBRA methodology. Afterwards, the application of CoBRA for the development of a GSD effort estimation model at Indra will be presented.

*A. Research Goal and Questions*

The main goal of this research was the development of an effort model for distributed and global software development projects. As we assume that there exists both a body of knowledge and established models for estimating the effort in collocated projects, we were specifically interested in the causes of effort overhead in GSD compared to collocated development. Thus, the research questions were:
1. What are the most important factors influencing overhead in a concrete GSD environment?
2. How are these factors interrelated?
3. How can the impact of the factors be quantified?

For answering these questions, we chose to develop a CoBRA model in one industrial GSD environment based on the experience of practitioners. The selection of CoBRA was done for specific reasons: First, not enough quantitative data is available for developing data-driven cost models on an empirical basis. Second, a large set of experiences is already available in GSD projects that can be collected and stored on a systematic basis. Thus, the effort estimations do not have to solely rely on individual expertise. CoBRA, as a hybrid approach for developing effort models using expert estimations and just little data, addresses these issues.

*B. CoBRA*

The Cost Estimation, Benchmarking, and Risk Analysis method (CoBRA) was developed by Fraunhofer IESE in 1996 and enhanced and applied in several studies [29-32]. Its main goal is to combine expert-driven and data-driven cost estimation methods in order to provide a method that can be used in organizations where only little data is available.

The basic formula used by CoBRA is:

*Cost = Nominal Productivity * Size + Cost Overhead.*

Thus, costs stem from two components: A nominal productivity determines the productivity in an optimal case – a project in which all cost drivers are at optimal values (e.g., the team is highly experienced, the requirements are of very high stability, and the customer very actively participates in the project). In CoBRA, it is assumed that there exists such a "best case" productivity within an organization that does not differ much between projects.

The cost overhead describes additional costs that occur in projects that are not done in an optimal environment. Thus, the overhead is driven by a set of influencing factors (e.g., team capability, requirements stability, or customer participation). These are organized in a causal model that determines their impact on project costs. Figure 1 gives an example of such a causal model.

The factors in a causal model can have (a positive or negative) impact in two ways:

Direct relationships describe a direct impact on cost overhead, which means that an increase of these factors will directly increase (respectively decrease for negative relationships) the cost overhead.

Indirect relationships determine the impact of another factor on cost overhead. In the example, increased requirements volatility increases cost overhead, but this increase is weakened if there is high discipline in requirements management.

Factors can have both a direct and indirect impact on cost overhead (as the disciplined requirements management in the example). Their impact on cost overhead is expressed in percentage of the nominal project costs, which makes the overhead also dependent on project size.

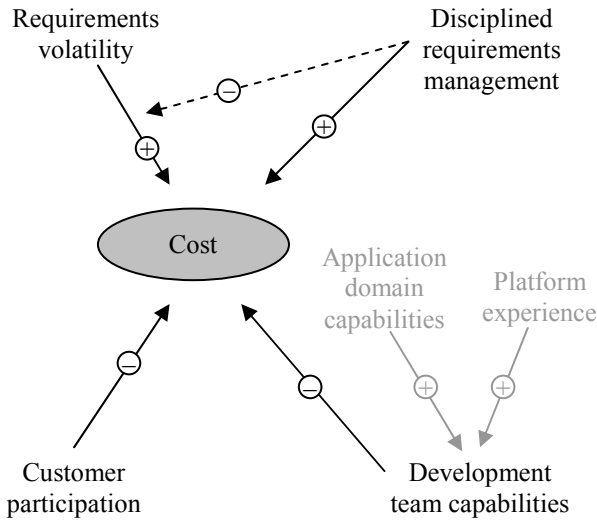

Figure 1. Causal model example [31]

In CoBRA, the development and quantification of the causal model is done by experts. Selection of the factors and determination of their causal relationships is done in group discussions. Afterwards, each expert estimates the quantitative impact of each factor on cost overhead. For direct relationships, the expert is asked to name the cost overhead caused by this factor in the worst case (e.g., customer participation is very low). For indirect relationships, the expert is asked to estimate the overhead in the worst case of the direct factor twice, for the best and worst case of the indirect factor. In order to reflect the inherent uncertainty, the estimation is not done as a single number but in a triangular distribution with the expert naming the minimum, maximum, and most likely overhead for each value.

Using the data from past projects, the nominal productivity is then computed. For this, the past projects are characterized in terms of the cost overhead influencing factors and the cost overhead is calculated using the quantified causal model. As only one variable needs to be determined, only few past projects are needed for this.

The CoBRA process can be summarized as follows:

1) Collect possible cost drivers: A set of possible factors influencing cost overhead is collected. This can be done in preparation of the CoBRA study but should be further refined in a group discussion with experts of the organization. It is suggested grouping the factors into categories such as product, personnel, project, and process.

2) Rank and select cost drivers: Each expert ranks the cost drivers individually according to their impact on cost overhead. Ranking can again be done within the factor categories. The aggregated results of the rankings are presented to the experts, and the most important cost drivers are selected in a group discussion.

3) Build causal model: In another group discussion among the experts, the direct and indirect influences of the factors on cost overhead are determined and a causal model is developed.

4) Quantify causal relationships: Based on the causal model, a questionnaire is prepared asking each expert to quantify the impact of the factors on cost overhead. In every judgment, each expert is asked to name the minimum, most likely, and maximum overhead in percent over a nominal (i.e., best) case.

5) Analyze past projects: For a set of past projects, size and effort data is collected and the experts are asked to characterize them with respect to the cost overhead drivers. Based on these results, the nominal baseline productivity is determined.

After these steps, the model can be used for estimating new software development projects.

### C. CoBRA Application for estimating GSD overhead at Indra

The CoBRA methodology was applied in the network of Software Labs of Indra (ISL). Indra is the premier IT company in Spain and a leading IT multinational in Europe. It employs around 28000 professionals, who work in different market segments such as Security & Defense, Transport & Traffic, and Energy & Industry. Based in Spain, Indra has over 35 subsidiaries worldwide and references in more than 100 countries.

ISL develops customized software solutions for Indra's markets. It has 20 development sites, half of which are located in Spain and the others in Latin America (Mexico, Panama, Colombia, Brazil, and Argentina), Slovakia, and the Philippines. Most of the software development projects at ISL are distributed – both within Spain and globally. Having initiated GSD due to the ability to work at customers' sites worldwide, having access to a large pool of human resources and reduced labor costs, ISL now practices distributed development successfully on a regular basis, but also experiences difficulties. Therefore, there is strong interest in analyzing the effort overhead in GSD and its causes.

Typically, ISL projects are organized into one leading site in Spain (which hosts project management and acts as an interface to the client) and one mirror site overseas. Sometimes, additional sites within Spain are involved that collaborate closely with the leading site. Thus, we used the concept of one main site and one mirror site as mental model for our research.

Six practitioners from ISL were interviewed for the model development. They had several years of experience in distributed development, with two being directors at ISL (each responsible for a business area), three being project managers, and one person working in the quality department.

The first three steps of the CoBRA process were conducted in a one-day meeting at the Indra facilities in Madrid. Even though the process followed the CoBRA methodology, several adaptations were made due to the different research goal and organizational constraints: Instead of asking for factors that influenced effort in general, we asked specifically for factors causing overhead in distributed development. In contrast to standard CoBRA, we thus asked the question:

*If you compare a distributed software development project to a project in which all work is done in Madrid, which factors cause the overhead that can be observed in distributed projects?*

Consequently, we did not use the original list and categorization of CoBRA as a starting point, rather but categorized the factors into "Project & Process", "Product", "Characteristics of the mirror site", and "Dependencies between sites". Table 1 shows the initial cost drivers we used as a basis for step 1. The drivers were collected based on the factors suggested by CoBRA, a literature research, and the results of previous empirical studies [23].

TABLE I. INITIAL SET OF EFFORT DRIVERS

| Project & Process | Product |
|---|---|
| • Requirements volatility<br>• Number of involved teams<br>• Customer acceptance of GSD<br>• Extent to which peer reviews and inspections are implemented<br>• Extent of disciplined requirements management<br>• Extent of disciplined configuration management<br>• Extend of having established communication procedures | • Importance of software quality<br>• Adequate documentation<br>• Importance of software maintainability<br>• Importance of software portability<br>• Importance of software reusability<br>• Complexity of the software |
| **Characteristics of mirror site** | **Dependencies between sites** |
| • Programmer capability<br>• Application experience<br>• Technical experience (language, tool, platform)<br>• Personnel continuity<br>• Willingness to cooperate<br>• Experience in working in a distributed way<br>• Communication ability with the customer site | • Language difference<br>• Cultural difference<br>• Time zone distance<br>• Infrastructure link<br>• Common experience of having worked together before<br>• Personal relationships and personal contacts between people at the different sites |

Due to the availability of the practitioners, the first three steps were not done in one group meeting but in a series of interviews. Each practitioner individually ranked the factors and formulated causal relationships. These results were then aggregated into one list of factors and a causal model. Based on the results, the questionnaire for step four was prepared and sent to the practitioners. All of the participants filled it out and returned it.

TABLE II. INDIVIDUAL RANKING OF DRIVERS

| 1: Director | 2: Quality department | 3: Project manager |
|---|---|---|
| **Site characteristics:**<br>• Process communication<br>• Application knowledge<br>**Dependencies:**<br>• Cultural differences<br>• Time zone differences<br>• Personal relationship<br>**Product:**<br>• Criticality<br>• Complexity<br>**Project & Process:**<br>• Importance of maintainability<br>• Time pressure | **Site characteristics:**<br>• Process knowledge<br>• Technical knowledge<br>• Project experience<br>**Dependencies:**<br>• Communication infrastructure<br>• Time zone difference<br>**Product:**<br>• Coupling between tasks<br>**Project & Process:**<br>• Peer reviews & inspections<br>• No of sites<br>• Configuration management | **Site characteristics:**<br>• Project experience (1)<br>• Process knowledge (2)<br>• Technical knowledge (3)<br>**Dependencies:**<br>• Common working experiences (1)<br>• Language difference (2)<br>• Personal relationships (3)<br>**Product:**<br>• Coupling between tasks (1)<br>• Formality of task description (2)<br>• Complexity (3)<br>**Project & Process:**<br>• Requirements stability (1)<br>• No of sites (2)<br>• Time pressure (3) |
| **4: Director** | **5: Project manager** | **6: Project manager** |
| **Site characteristics:**<br>• Personnel continuity<br>• Transparency<br>• Process knowledge<br>• Staff motivation<br>**Dependencies:**<br>• Language difference<br>• Cultural difference<br>• Communication infrastructure<br>• Common working experiences<br>• Organizational & goals alignment<br>**Product:**<br>• Criticality<br>• Complexity<br>• Formality of description<br>• Coupling to other tasks<br>**Project & Process:**<br>• Requirements stability<br>• No of sites<br>• Customer acceptance<br>• Disciplined configuration management | **Site characteristics:**<br>• Communication ability with customer (1)<br>• Transparency (2)<br>• Technical knowledge (3)<br>**Dependencies:**<br>• Communication infrastructure (1)<br>• Personal relationships (2)<br>• Common working experience (3)<br>**Product:**<br>• Criticality (1)<br>• Complexity (2)<br>• Formality of task description (3)<br>**Project & Process:**<br>• Requirements stability (1)<br>• Extent of disciplined requirements management (2)<br>• Architecture stability (3) | **Site characteristics:**<br>• Staff motivation (1)<br>• Project experience (2)<br>• Communication ability with customer sites (3)<br>**Dependencies:**<br>• Language difference (1)<br>• Cultural difference (2)<br>• Personal relationships (3)<br>**Product:**<br>• Complexity (1)<br>• Importance of SW maintainability (2)<br>• Importance of SW reusability (3)<br>**Project & Process:**<br>• Requirements stability (1)<br>• Extent of disciplined requirements management (2)<br>• Extent of disciplined configuration management (3) |

Performing individual interviews instead of a group meeting might lead to large differences in the answers and thus might make it hard to integrate them into one model. However, we found many similarities in the rankings and therefore do not see this as a threat to validity. Instead, this could be seen as a reduction of the bias that might be created in a group meeting where the result might be dependant on social interactions between the group members (e.g., one person dominating the group).

Unfortunately, not enough project effort and size data was available at ISL: As the causal model specifically contained overhead drivers of GSD, other effort drivers were disregarded. Thus, the analyzed projects had to be very homogenous with respect to the disregarded drivers in order to provide a common effort baseline. However, it was not possible to access such a set of projects during the study. Therefore, step five could not be completed and no calculation of the baseline productivity and validation of the model could be done yet. This remains future work and is planned to be done in a later study.

## IV. RESULTS

### A. Findings

Table 2 gives an overview of the results of the individual rankings. As some of the interviewees did not rank the factors from highest to lowest but instead just marked the most important factors for each category, we present the answers of all practitioners as a list of most important factors (while considering the three highest-ranked factors in each category as most important) and name the ranks in parentheses if the factors were ranked.

Table 3 shows the results of the identified causal relationships (interviewees 4 and 5 identified the causal relationships together in one interview session.). Compared to the factor rankings, the interviewees' answers regarding the causal relationships differed to a large extent. However, some commonalities could still be identified.

In order to aggregate the individual estimations into one coherent model, we used the following rules: (1) All factors that were named by at least three interviewees as one of the most important factors became factors of the model and (2) all causal relationships that were named at least twice were used to build the causal model. Due to the fact that many interviewees refused to rank the factors (and only picked the most important ones without ranking them), more sophisticated statistical methods could not be used here.

Two adaptations were made to the results: Most of the interviewees named at least one of the factors "process knowledge", "transparency", or "communication ability" describing the remote site as one of the most important factors. However, they all seemed to mean similar things with these factors. We thus summarized them under the term "process maturity", meaning the extent to which defined processes for handling work, description and documentation of results, and communication are known and are followed. The factors "language differences" and "cultural differences" were named by several of the practitioners as influencing all other factors. As the CoBRA model gets very complicated if several factors have an indirect influence on one other factor and as the two factors described very similar phenomena, we summarized them under "language and cultural differences".

TABLE III. INDIVIDUAL CAUSAL RELATIONSHIPS

| |
|---|
| **1: Director**<br>• Time pressure → Process communication<br>• Personal relationship → Process communication |
| **2: Quality department**<br>• Process knowledge → Peer reviews & inspections<br>• Process knowledge → Coupling between tasks |
| **3: Project manager**<br>• Project experience → Technical knowledge<br>• Language → Personal relationships<br>• Personal relationships → Coupling |
| **4 & 5: Director & Project manager**<br>• Personnel continuity → Technical knowledge<br>• Stability of requirements → Technical knowledge<br>• Transparency → Organizational alignment<br>• Language / cultural differences → Organizational alignment<br>• Coupling between tasks → Organizational alignment<br>• Coupling between tasks → Configuration management<br>• Criticality ←→ complexity<br>• Language / cultural differences → all other factors<br>• Transparency / communication → all other factors |
| **6: Project manager**<br>• Language, cultural differences, personal relationships, transparency, communication → all other factors<br>• Requirements stability → Reusability, maintainability<br>• Characteristics of site → Traceability, maintainability |

Figure 2 shows the resulting causal model. It shows that ultimately, 14 factors were selected with 12 of them directly influencing effort overhead and 2 having indirect influences. Out of the 14 factors, there were 3 characteristics of sites, 4 of tasks, 3 of project and process, and 4 relationships between sites.

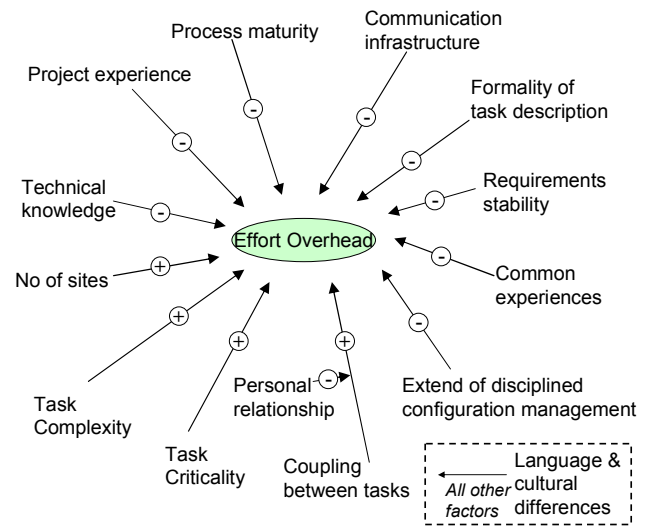

Figure 2. Resulting causal model

Using this causal model, the questionnaire for model quantification was created and answered by 4 of the participants. From the resulting estimations, the average values were created, giving an aggregated estimation on the quantitative impact of each factor on effort overhead. Each of the estimations was given in three values (minimum, most likely, maximum). In order to avoid excessive complexity, we will only present the average numbers for the estimated "most likely" values. Table 4 gives an overview of the aggregated results. As explained in Section 3B, the estimations always cover the "worst case" of the influencing factor and describe the overhead for a nominal project under optimal conditions.

For 11 of the direct influencing factors, two values are given: The causal model indicates that these factors are influenced by the factor "language and cultural differences". Consequently, there exists one value for the case that language and cultural differences are very high and an additional estimation for the case that these differences are very low. The coupling between tasks also depends on personal relationships and thus requires four estimations for the combinations of very high and very low personal relationships with language and cultural differences.

Table 4 also shows the standard deviations in the estimations for each value (in parentheses). It can be seen that process maturity, formality of the task description, and requirements stability are seen as the effort drivers with the highest impact – however, with different sensibilities towards language and cultural differences: While the impact on task description formality is rather small (only 10% distance between high and low cultural and language differences), the overhead of a low process maturity is very much dependent on language and cultural differences (nearly 20% distance).

TABLE IV. RESULTS OF THE MODEL QUANTIFICATION

| Factor | | Very low language & cultural differences | Very high language & cultural differences |
|---|---|---|---|
| Process maturity (Site) | | 26.25% (18) | 45.00% (23) |
| Project experience (Site) | | 15.50% (12) | 36.25% (32) |
| Technical knowledge (Site) | | 18.75% (8) | 33.75% (15) |
| Communication infrastructure (Site dep.) | | 18.75% (7) | 30.00% (18) |
| Common experiences (Site dep.) | | 9.25% (9) | 16.50% (14) |
| Criticality (Task) | | 16.25% (10) | 28.75% (18) |
| Complexity (Task) | | 17.50% (10) | 32.50% (22) |
| Formality of description (Task) | | 28.75% (27) | 38.75% (28) |
| Requirements stability (Project) | | 29.25% (20) | 42.75% (27) |
| Number of sites (Project) | | 10.50% (11) | 19.00% (18) |
| Disciplined configuration management (Project) | | 19.75% (18) | 27.75% (25) |
| Coupling (Task) | High relationships | 9.50% (7) | 16.25% (5) |
| | Low relationships | 20.75% (13) | 32.00% (18) |

## B. Threats to validity

The findings of this study face several threats to validity, which we will briefly address in the following. We categorize them into internal (are the observed phenomena based on a cause-effect relationship?), conclusion (are the results statistically significant?), construct (do the measures reflect the real world?), and external (can the findings be generalized?) threats:

Internal validity might be threatened by interviewees not being willing to correctly state their personal estimations. However, as all interviewees were very open, we see this threat as rather small.

Conclusion validity is threatened by the relatively low number of respondents, which does not allow for making statements with high statistical significance. This is also shown by the high standard deviation in most of the quantifications. Therefore, we are planning to conduct similar studies with a larger number of estimations and data sources.

Construct validity might be threatened by the interviewees having different understandings of effort overhead and the individual cost drivers. We tried to minimize this threat by giving detailed definitions of every cost factor.

External validity is threatened by the fact that all interviews were done within only one organization. It is very likely that in other organizations, the relative weights and the quantifications of the factors will look differently. However, we think these results can provide a general overview of the importance of different cost drivers in GSD that is also applicable to other environments.

## V. TOWARDS WORK DISTRIBUTION EVALUATION

In the following, we will briefly sketch how the results of a causal model and a quantification of its factors can be used in a model for quantitatively evaluating different task allocation scenarios and can thus be used to support work distribution in GSD.

In task allocation, there is a significant difference between effort and development costs: As cost rates may be different at different sites, a certain assignment alternative (e.g., assigning much work to a low-cost site) might lead to *higher* effort but *lower* development costs. Thus, the effort at each site has to be multiplied with the site-specific cost rate in order to get an estimation of overall development costs [8].

In an earlier publication [33], we defined a process for evaluating task allocations that is based on a CoBRA model for effort estimation. In contrast to the mental model used here (one main site and one mirror site), we assumed a scenario consisting of several distributed sites and a set of defined tasks that is to be distributed among the sites. This means that the factors might have different values within one project: Site factors (e.g., project experience) can be different for every site, site relations (e.g., common experiences) for every combination of two sites, and task characteristics (e.g., criticality) for each of the tasks that are to be distributed. However, the factors, causal relationships, and quantifications identified here could still be used in this more complex environment.

The process for task allocation evaluation is based on the Goal-Question-Metric paradigm [34] and done as follows [33]:

1. The decision maker (i.e. the person responsible for task allocation) who defines the viewpoint for the evaluation is identified.
2. The project context (i.e., the tasks and sites together with their characteristics) is identified.
3. The evaluation criteria for analyzing task assignments are selected. Typically, one of the most important criteria is the development cost for which the expected effort has to be estimated.
4. The factors that have an allocation-dependent influence on the evaluation criteria are identified. For the criteria effort and cost, these factors are the ones identified in this study.
5. The baseline values for the task allocation evaluation criteria are identified. With respect to the effort estimation, the baseline effort is the effort for a collocated project that is increased by the overhead due to distribution. The baseline effort can be obtained using historical data, expert estimations, or cost models such as COCOMO [2].
6. The impact of the influencing factors on the evaluation criteria is identified. For the effort and cost criteria, this impact is represented by the causal model and its quantification as developed in this study.
7. For the identified factors, the corresponding values for every site (site characteristics), every pair of sites (site dependencies), and every task (task characteristics) are assessed and documented. This allows for evaluating the different task distribution possibilities.
8. As now a CoBRA-based model is established and the values for the different tasks and sites have been assessed, the different task assignment alternatives can be evaluated: For every specific assignment, the values of the corresponding tasks and sites are inserted into the effort overhead model and (together with the identified baseline effort), the estimated effort for this assignment can be calculated and compared to the effort predictions of other assignments. For evaluation criteria other than effort, the evaluation is done in a similar way. If, for example, development costs were selected as a criterion, the estimated effort for each site can be multiplied with the site-specific cost rate (similarly to the approach by Madachy [8]).

Finally, this process results in systematic, quantified evaluations of different task assignment alternatives and thus can help to find the optimal task assignment individually for a specific project.

In [33], we demonstrated how this process can be instantiated towards quantitatively evaluating different assignment scenarios with respect to the expected effort and development costs. In a hypothetical scenario, we described the decision of transferring work to a newly established site in India made by a European development company. For this scenario, we developed a quantified effort model, which was implemented as a simple Excel sheet and incorporated the variation factor selection, the impact of the factors on productivity, and the evaluation of assignment scenarios (steps 4 and 6-8). Figure 3 gives an impression of the developed Excel model.

However, the factor selection, the identification of their causal relationships, and the quantification of the impact were done only hypothetically in this example, based on our own estimations. The process demonstrated in this paper and the identified results represent a way for gathering this information empirically and for systematically reusing the experiences of experts. Therefore, the combination of the two approaches can be used for assessing and selecting task assignments with respect to an organization-specific evaluation of the expected effort and costs for every assignment alternative.

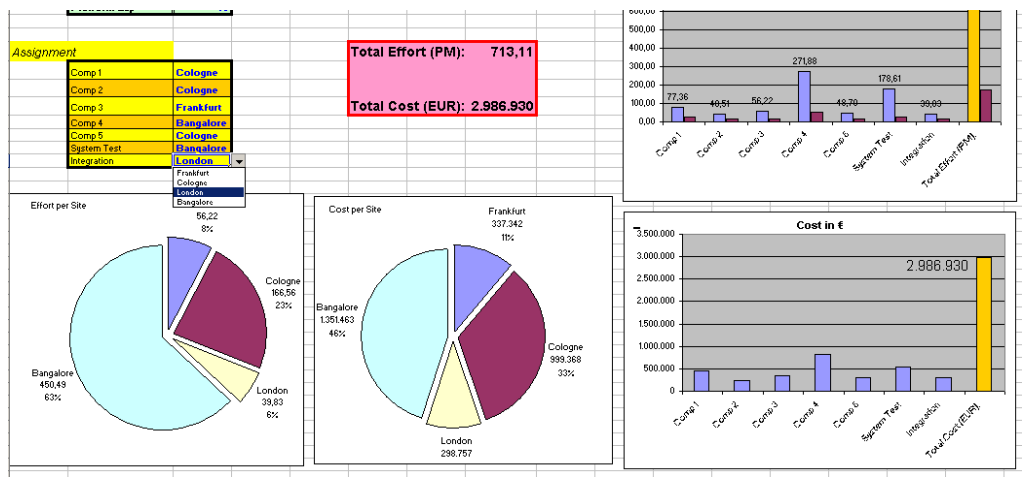

Figure 3. Excel model for evaluating task assignments with respect to effort and cost

## VI. CONCLUSION AND FUTURE WORK

In this article, we developed a model for estimating the overhead in distributed development projects based on a set of influencing factors, their causal relationships, and a quantification of their impact on effort overhead. The results of this study can be used twofold:

First, the identified factors and their estimated impact on productivity can be used by other organizations for getting an overview of the most import effort drivers in distributed development. This can help to assess the overhead in GSD and thus to get a realistic idea of the effects of distribution on productivity. Therefore, the list can help to overcome the problem of only considering the possible labor cost savings of GSD without seeing them in combination with the likely productivity downfall. In addition, the list of factors can be used for identifying possibilities for improving productivity in distributed development: As these factors have the highest impact on productivity, improving them would presumably provide the best leverage for increasing productivity in GSD.

Second, the model development process presented here can be applied by other organizations in order to develop their own effort overhead models. This can help organizations in effectively predicting their effort overhead individually for new projects and thereby improve their project planning and task allocation in GSD.

However, the model and results presented here also possess some limitations that should not be overlooked:

The results represent only one specific environment and cannot be easily generalized to other contexts and organizations. The individual rankings of the factors and especially the quantitative impact on productivity overhead are organization-specific and might most probably look very different in any other environment. Therefore, the model should not be used as is in other effort prediction approaches for GSD. However, we think that the list of factors, the causal model, and the relative weights can generally be indicators for important effort drivers in GSD and their relationships: Many problems described by the managers in the interviews (e.g., misunderstandings, cultural clashes, and problems with low process maturity) are reported repeatedly in various distributed development projects and can therefore be assumed to exist in almost every GSD environment. Thus, their impact factors will probably also be similar in a different environment – which indicates that the results identified here are likely to be applicable to many different organizations. However, more empirical studies are needed to validate this hypothesis.

Additionally, the quantification of the factors' impact on productivity presumes a high degree of objectivity and precision. However, concepts such as the impact of cultural differences on productivity are very hard to quantify and the results should be treated with care. This is due to the complexity and unpredictability in human behavior, which – especially in distributed development – has the highest impact on the resulting effort overhead.

However, the use of a quantitative model for effort overhead prediction still entails many benefits: All large software development projects require some sort of effort estimation, which always has to deal with the uncertainty in predicting human behavior. Using such a model can help to make the underlying assumptions and estimations explicit and lead towards a systematic process of analyzing and predicting effort in GSD. In addition, the model can be the basis for reusing experiences and lessons learned in distributed projects by updating the underlying causal relationships and quantifications based on project results. Finally, the process of developing a model helps to understand the characteristics and problems of distributed development, which might increase the project planning skills of the people involved.

In the future, we aim at integrating the effort overhead model with other models for evaluating and identifying task assignments that are able to cope with the inherent uncertainty better. For example, we developed a task distribution model based on Bayesian networks that specifically addresses prediction uncertainty [35] and are developing risk models that assess task distributions under the perspective of possible project risks. In future work, we will integrate these different models into one approach for systematic task assignment that is able to evaluate task assignments under various perspectives.

Additional future work will be the refinement of the model presented here and its quantification. This will be done in accordance with the CoBRA process based on a characterization of historical projects.

### ACKNOWLEDGMENT

The authors would like to thank all participants in the interview studies. At ISL, these persons were: Angel Villodre, Pablo Jesus Sanchez Moreno, Julian Diaz del Campo Jimenez de los Galanes, Francisco Fernández Fabián, Carlos Alger López, Diego Jiménez Romero, Ramón Torres, Ana López Díaz, David Graña, Ana Gómez-Escolar, Jose Luis Moragas, Manuel Cerrillo, Javier F. Gómez, Cindy Pinato, and Paloma Martínez Tordesillas. The authors also thank Sonnhild Namingha for proofreading the paper.